\documentclass[amssymb,twocolumn,prd,superscriptaddress]{revtex4-1}

\usepackage{amsmath}

\newcommand{\be}{\begin{equation}}
\newcommand{\ee}{\end{equation}}

\begin{document}

\title{TeV Scale  Cross-Sections  and the Pomeranchuck Singularity}

\author{G. C. Joshi}

\affiliation{School Of Physics, University Of Melbourne, Parkville, Victoria 3010, Australia}

\begin{abstract} 
We have investigated the detailed structure of
$l$-plane singularities of scattering amplitude saturating the
Froissart bound. A self-consistent analysis of these singularities
provides us secondary terms in the Froissart bound. These secondary
terms lead to ghosts in the $l$-plane, which can only be removed by
introducing an odderon singularity. Phenomenological implications of
this analysis are also discussed.
\end{abstract}

\maketitle

TOTEM and LHCf experiments at LHC \cite{1,2} have revived new interest
in the high energy behaviour of scattering cross-sections. Measurement
of high-energy cross-sections at energies $\sqrt{s}$ = 14 TeV will
provide a deep insight into the dynamics of hadronic interactions and
some of the most important principles of physics.

At present our theoretical understanding of physics at these energies
is rather incomplete. There are a number of theories like soft QCD,
eikonal and most important, the Regge theory. This theory has a
remarkable history in explaining high-energy behaviour in terms of few
parameters. Furthermore, with dual models like Veneziano
representation, these theories provide a unified description of
high-energy behaviour and low energy resonances.

In this paper we will investigate the high-energy behaviour of the
Pomeranchuck singularity based on the most general principles of
physics:

\begin{itemize}
\item[(i)] Unitarity and
\item[(ii)] Analyticity.
\end{itemize}

Our starting points will be the Froissart bound \cite{3} and the
one-dimensional dispersion relations.  We will first calculate the
l-plane singularities using the Froissart bound. Then from these
singularities we will derive the high-energy behaviour of the
scattering amplitude. We will show that there is an interesting
relationship between the detailed structure of the l-plane
singularities and the detailed structure of the high-energy
behaviour. In the language of the 60's, we are going to ``bootstrap''
the Pomeron.

For simplicity we start with spin zero kinematics where the t-channel
Froissart-Gribov \cite{4} representation for the partial wave
amplitude is given by

\be
a_l(t) = \frac{2}{t-4m^2} \int_{4m^2}^{\infty} Q_l \left(1+\frac{2s}{t-4m^2}\right) A_s(t,s) ds
\ee

This representation via Carlson's theorem \cite{5} provides a unique
interpolation to the complex angular momentum plane. In the 60's
several authors \cite{3} including this author \cite{6,7,8}
investigated the analytic properties of the partial wave amplitude
$a_l(t)$ in the complex angular momentum plane and showed that
$a_l(t)$ was a meromorphic function with moving poles at
$l=\alpha(t)$.

In this paper we will assume Froissart bound for $A_s(t,s)$

\be
\label{eq:2}
A_s(t,s) \leq \beta(t) \log^2\left(\frac{s}{s_0}\right) \; {\rm for} \; s\geq N
\ee

where $N$ is a large number. We can now write

\be
a_l(t) = A(l,t) + B(l,t)
\ee

where

\be
\begin{split}
\label{eq:4}
&A(l,t) = \frac{2}{t-4m^2} \times\\
&\int_N^{\infty} Q_l\left( 1 + \frac{2s}{t-4m^2}\right) \beta(t) s^{\alpha(t)} \log^2\left(\frac{s}{s_0}\right) ds
\end{split}
\ee

and

\be
\begin{split}
&B(l,t) = \frac{2}{t-4m^2}\times\\
& \int_{4m^2}^N Q_l\left( 1 + \frac{2s}{t-4m^2} \right) A_s(t,s) ds
\end{split}
\ee

Here $\alpha(t)$ is the Pomeranchuck trajectory with $\alpha(0) = 1$
and $\mathcal{R}e \alpha(t) \leq 1$. Detailed structure of $\alpha(t)$
will be discussed later on.

In eq.~(\ref{eq:4}) expanding $Q_l(z)$ for large $z$ we obtain

\be
\begin{split}
\label{eq:6}
A(l,t) &= 2^{-1-2l} \sqrt{\pi} (t-4m^2)^l \frac{\Gamma(l+1)}{\Gamma(l+3/2)} \beta(t) s_0^{-2\alpha(t)} \times\\
& N^{\alpha(t) - l} \Big\{ \frac{1}{l-\alpha(t)} \log^2\left(\frac{N}{s_0}\right) \\
&\quad\quad- \frac{2}{[l-\alpha(t)]^2} \log\left(\frac{N}{s_0}\right) \\
&\quad\quad+ \frac{2}{[l-\alpha(t)]^3}\Big\}
\end{split}
\ee

The above integration is performed in the domain $l>\alpha(t)$. The
resulting representation eq.~(\ref{eq:6}) now provides an analytic
continuation of $A(l,t)$ in the entire $l$-plane with simple, double
and triple moving poles at $l=\alpha(t)$.  $A(l,t)$ also has the usual
fixed poles at $l=-1,-2,\ldots$. For the other part $B(l,t)$ we expand
$A_s(t,s)$ in a Taylor series

\be
A_s(t,s) = \sum_n c_n(t) \left(\frac{s}{s_0}\right)^n  \frac{t-4m^2}{2}
\ee

and use representation of $Q_l(z)$ at $z\sim1$ to obtain

\be
\begin{split}
\label{eq:8}
&B(l,t) = \sum_n c_n(t) \Biggr\{ \frac{(N/s_0)^{n+1}}{n+1}\times\\
& \left[ \log\left(\frac{N}{s_0}\right) - 2 - \gamma - \psi(l+1) \right]\\
&-\frac{(4m^2/s_0)^{n+1}}{n+1}\times\\
& \left[ \log\left(\frac{4m^2}{s_0}\right) - 1 + \log\left(\frac{t-4m^2}{s_0}\right) - \gamma - \psi(l+1)\right] \Biggr\}
\end{split}
\ee

where $\psi(z) = \Gamma'(z)/\Gamma(z)$ and $\gamma$ is Euler's
constant. This representation $B(l,t)$ is an analytic function except
for fixed poles at $l=-1, -2, \ldots$.

Now using the singularities of eq.~(\ref{eq:4}) in terms of a single,
double and a triple pole we can calculate the asymptotic behaviour of
$A(s,t)$ via the Sommerfeld-Watson transform i.e.

\be
\label{eq:9}
A(s,t) = \frac{1}{2i} \int_c (2l+1) \frac{a(l,t)}{s_{m\pi l}} P_l(-z) dz
\ee

where the contour is clock-wise and the signature factor is included
in the definition of $a(l,t)$.

Taking the residue of poles in eq.(\ref{eq:9}) we get

\be
\begin{split}
A(s,t) &= XYP \log^2 \left(\frac{N}{s_0}\right)\\
& - 2 [ X'YP + X (Y'P + P'Y)] \log\left(\frac{N}{s_0}\right)\\
& + X'' Y P + 2 X' (Y'P + P'Y) \\
&+ X(Y'' P + Y' P' + P'' Y + P'Y')
\end{split}
\ee

evaluated at $l=\alpha(t)$, where primes denote differential with
respect to $l$ and evaluated at $l=\alpha(t)$ and

\be
\begin{split}
X(l,t) &= \frac{2^{-1-2l}}{2l+1} \sqrt{\pi} (t-4m^2)^l \times\\
&\frac{\Gamma(l+1)}{\Gamma(l+3/2)} \beta(t) s_0^{-2\alpha(t)} N^{\alpha(t)-l}
\end{split}
\ee

\begin{subequations}
\be
Y(l,t) = - \frac{1 + e^{-i\pi l}}{s_{m\pi l}}
\ee
\be
P=P_l(-z) = P_l\left(-1-\frac{2s}{t-4m^2} \right)
\ee
\end{subequations}

At this point the high energy behaviour of of the scattering amplitude
as given by eq.(\ref{eq:8}) has a pathology.  We call this ``odderon
anomaly''.  This comes from the $X Y'' P$ term in eq.(\ref{eq:8})
which in its full form can be written as

\be
\begin{split}
\label{eq:13}
&-X P \Big[ -\frac{\pi^2}{2} \left(\frac{1+e^{-i\pi\alpha(t)}}{\sin\pi\alpha(t)}\right) - \frac{\pi^2}{2} \left(\frac{1-e^{-i\pi\alpha(t)}}{\sin\pi\alpha(t)}\right)\\
&+2i\pi^2 \frac{e^{-i\pi\alpha(t)}}{\sin^2 \pi\alpha(t)} \cos\pi\alpha(t)\\
& + 2\pi^2 \frac{1+e^{-i\pi\alpha(t)}}{\sin^3\pi\alpha(t)} \cos^2\pi\alpha(t)\Big].
\end{split}
\ee

The first term in eq.~(\ref{eq:13}) is the usual Pomeron term with
positive signature.  And the second term is the odd-signature Pomeron
(odderon). All terms in eq.~(\ref{eq:13}) are well behaved at
$\alpha(0)=1$ except the odderon term, which has a ghost. Conventional
ghost killing mechanisms like the Chew mechanism \cite{9} or Gell-Mann
mechanism \cite{10} do not work here.  The basic idea behind
ghost-killing mechanism is that when $\alpha(0)=1$ the pole residue
develops a zero at this point removing the ghost. This idea cannot
work here because if $X$ develops a zero at $\alpha(0)=1$ a large
number of of terms also vanish because they also have the same
residue. This also removes the most important term

\be
X Y P'' \sim s^{\alpha(t)} \log^2 \left(\frac{s}{s_0} \right)
\ee

which is our assumption regarding the asymptotic behaviour.

However, we can remove the ghost by introducing an additional term in
eq.~(\ref{eq:2}) i.e

\be
-X P \frac{\pi^2}{2} \left(\frac{1-e^{-i\pi\alpha(t)}}{\sin\pi\alpha(t)} \right)\frac{\gamma(t)}{\beta(t)}
\ee

such that 

\be
\gamma(0) = \beta(0)
\ee

With this new term we can recalculate the singularities and then using
the Sommerfeld-Watson \cite{3} transform we get

\be
\begin{split}
&A(s,t) = \Big\{ X Y T \log^2 \left(\frac{N}{s_0}\right) \\
&- 2[X' Y T + X(Y' T + T' Y)] \log\left(\frac{N}{s_0}\right)\\
& +X''Y T + 2 X' Y' T + 2 X' Y T' \\
&+ X (Y'' T + 2 Y' T' + Y T'')\Big\}(-z)^{\alpha(t)}\\
&+\Big[-2X Y T' \log\left(\frac{N}{s_0}\right) + 2 X' Y T \\
&+ 2 X Y' T + 2 X Y T'\Big] (-z)^{\alpha(t)} \log(-z)\\
&+ X Y T (-z)^{\alpha(t)} \log^2 (-z)
\end{split}
\ee

where

\be
P_l(_z) = \sqrt{\pi} \frac{\Gamma(l+1/2)}{\Gamma(l+1)} (-z)^l = T(l) (-z)^l
\ee

evaluated at $l=\alpha(t)$ and

\begin{subequations}
\be
T' = \left[ \frac{\partial T(l)}{\partial l}\right]_{l=\alpha(t)}
\ee
\be
T'' = \left[ \frac{\partial^2 T(l)}{\partial l^2}\right]_{l=\alpha(t)}.
\ee
\end{subequations}

Thus using the optical theorem which is also based on unitarity

\be
\nonumber
\sigma_{\rm TOT} = \frac{8\pi}{q_s\sqrt{s}} {\rm Im}[ A(s,t=0) ]
\ee

 the total cross-section can be written as

\be
\begin{split}
\label{eq:19}
\sigma_{\rm TOT} &= 16\pi\Bigg\{ M[\alpha(0)] + N[\alpha(0)] \log\left(\frac{s}{2m^2}\right) \\
&+ X Y T \log^2\left(\frac{s}{2m^2}\right)\Bigg\}
\end{split}
\ee

with

\begin{subequations}
\be
\begin{split}
M &= X Y T \log^2 \left(\frac{N}{s_0}\right) \\
&- 2[X' Y T + X (Y'T + Y')] \log\left(\frac{N}{s_0}\right)\\
& + X'' Y T + 2 X' Y' T + 2 X' Y T'\\
&+X (Y''T + 2 Y' T' + Y T'')
\end{split}
\ee
and
\be
\begin{split}
N &= -2 X Y T \log\left(\frac{N}{s_0}\right) + 2 X' Y T \\
&+ 2 X Y' T + 2 X Y T'.
\end{split}
\ee
\end{subequations}

Where we have used the trajectory $\alpha(t) = \alpha(0) + \alpha' t =
1+\alpha' t$.

We note that in eq.~(\ref{eq:19}) there is no odderon contribution,
however for small values of $t$ both Pomeron and odderon will
contribute. We also note that factorisation property
\cite{11,12,13,14,15} will not hold for eq.~(\ref{eq:19}).

We will now discuss nature of Pomeranchuck trajectory. For our
analysis all we need to assume is that for Pomeron $Re \alpha(t)\leq1$
and $\alpha(0)=1$. There are several examples of such trajectories
like

\be
\alpha(t) = 1-\alpha' \sqrt{-t}
\ee

and

\be
\label{eq:alphalog}
\alpha(t) = 1-\alpha_1 \log(1+\alpha_2 t^2).
\ee

It should be noted that this parameterization is valid only near
$t\sim0$.  The p-p total cross-section will also get a contribution
from secondary trajectories. From the point of view of duality there
are three Veneziano amplitudes $V(s,t)$, $V(s,u)$ and $V(t,u)$. As
s-channel is exotic only $V(t,u)$ will contribute. Here there are two
types of mesons normal (Q$\bar{{\rm Q}}$) trajectories like $\rho-A_2$
and the baryonium trajectories (QQ$\bar{{\rm Q}}\bar{{\rm Q}}$)
\cite{19,20,21,22,23,24}. As so far no baryonium are found one expects
baryonium trajectories will have a smaller slope compared to
(Q$\bar{{\rm Q}}$) trajectories . Thus such trajectories will only
contribute for large $t$.

A detailed phenomenological analysis of $p-p(\bar{p}-p)$ and
$\pi^{\pm}-p$ have been carried out by several authors. Donnache
\cite{25} has used a form

\be
\nonumber
\sigma_{\rm TOT} = X s^{-\epsilon} + Y s^{-\eta}
\ee

where the first term is the Pomeron contribution and the second term
is conventional $\rho-A_2$ trajectories.

Block and Halzen \cite{25} have used a form in terms of lab energy
$\nu$

\be
\sigma^{\pm} = c_0 + c_1 \log\left(\frac{\nu}{m}\right) + \beta_{p'} \left(\frac{\nu}{m}\right)^{\mu-1} \pm \delta\left(\frac{\nu}{m}\right)^{\alpha-1}.
\ee

These authors make a consistent fit these cross-sections.  Both these
forms can be obtained from our eq.~(\ref{eq:alphalog}).

This work was supported by the Australian Research Council.  I would
like to thank Bruce McKellar, Geoff Taylor, Ray Volkas, Robert Foot
and Archil Kobakhidze for interesting discussions and encouragement.


\begin{thebibliography}{25}

\bibitem{1} Proceedings of MPI@LHC'08, Perugia, Italy, 27- 31 October, 2008. Proceedings of the Elastic and Diffractive Scattering Workshop 
(Blois 2007) DESY, Hamburg, Germany, May 2007. 

\bibitem{2} F. Ferro, TOTEM Experiment at the LHC: Status and Program International Workshop on Diffraction in High-Energy Physics, Adamantas, Greece, 5 - 10 Sep 2006, pp. 019.

\bibitem{3} M. Froissart, Phy. Rev., 122, 103 (1961).

\bibitem{4} see for example E. J. Squres, ``Complex Angular Momentum and Particle Physics'' published by W.A. Benjamin, Inc New York (1963).

\bibitem{5} see for example E. C. Titchmarsh,  ``Theory of Functions'' Oxford University Press. New York (1939).

\bibitem{6} G. C. Joshi and H. Banerjee, Phys. Rev., 137, B1567, (1965).

\bibitem{7} G. C. Joshi, Phys. Rev., 141,  1471, (1966).

\bibitem{8} G. C. Joshi, Nuovo Cimento, XLA:  630, (1965).

\bibitem{9} G. F. Chew, Phys. Rev. Letters, 16, 60, (1966).

\bibitem{10} M. Gell-Mann, in Proceedings of the International Conference on High Energy Nuclear Physics, Geneva 1962 edited by J.Prentki (CERN), p.539 (1962).

\bibitem{11} G. C. Joshi Nuovo Cimento 10, 821 (1975); 13, 680, (1975).

\bibitem{12} G. C. Joshi and S.Y. Lo, Nucl. Phys. B 93,  405, (1975).

\bibitem{13} G. C. Joshi, S. Y. Lo and B. Kellett, Nuovo Cimento 41A, 331, (1977).

\bibitem{14} G. C. Joshi, S. Y. Lo and B. Kellett, Nuovo Cimento 47A, 281, (1979).

\bibitem{15} G. C. Joshi, S. Y. Lo and B. Kellett, Nuovo Cimento 41A, 351, (1977).

\bibitem{16} G. C. Joshi, A. W. Martin, Physical Review, 188: 2354, (1969).

\bibitem{17} G. C. Joshi and A. Pagnamenta, Phys. Rev. D 1, 3117, (1970).

\bibitem{18} G. C. Joshi and A. Pagnamenta, Particles and Fields 1,  220, (1970).

\bibitem{19} G. C. Joshi, R. Anderson, J. Math. Phys., 20, 1015, (1979).

\bibitem{20} G. C. Joshi and R. Warner, Hadronic Journal, 2, 198, (1979).

\bibitem{21} G. C. Joshi, R. Anderson, Phys. Rev., 20,  736, (1979);
Phys. Rev., 20, 1666, (1979).

\bibitem{22} G. C. Joshi and R. Warner,  Phys. Rev. 22:  1012, (1980).

\bibitem{23} R. Ellis, R. Anderson, G. C. Joshi and B. H. J. McKellar 
Phys. Rev. D, 22, 2832, (1980).

\bibitem{24} R. Ellis, B. H. J. McKellar and G. C. Joshi, Nuovo Cimento Letts., 30, 355, (1981).

\bibitem{25} A. Donnachie, CERN preprint CERN- TH6635/92;
 M. M. Block and F. Halzen, Physical Review D 72, 036006, (2005).

\end{thebibliography}
\end{document}